\documentstyle[epsfig,12pt]{article}
\newcommand{\be}{\begin{eqnarray}}
\newcommand{\ee}{\end{eqnarray}}

\def\lsim{\mathrel{\rlap{\lower4pt\hbox{\hskip1pt$\sim$}}
\raise1pt\hbox{$<$}}}               
\def\gsim{\mathrel{\rlap{\lower4pt\hbox{\hskip1pt$\sim$}}
\raise1pt\hbox{$>$}}}               

\begin{document}

\rightline{\Large{Preprint RM3-TH/99-2}}

\vspace{2cm}

\begin{center}

\Large{Relativistic effects on the neutron charge form factor\\[3mm] in the
constituent quark model\footnote{\bf To appear in Physics Letters B.}}\\

\vspace{2cm}

\large{Fabio Cardarelli and Silvano Simula}\\

\vspace{1cm}

\normalsize{Istituto Nazionale di Fisica Nucleare, Sezione Roma III,\\ Via
della Vasca Navale 84, I-00146 Roma, Italy}\\

\end{center}

\vspace{1cm}

\begin{abstract}

\noindent The neutron charge form factor $G_E^n(Q^2)$ is investigated within
a constituent quark model formulated on the light-front. It is shown that,
if the quark initial motion is neglected in the Melosh rotations, the Dirac
neutron form factor $F_1^n(Q^2)$ receives a relativistic correction  which
cancels exactly against the Foldy term in $G_E^n(Q^2)$, as it has been 
recently argued by Isgur. Moreover, at the same level of approximation the 
ratio of the proton to neutron magnetic form factors $G_M^p(Q^2) /
G_M^n(Q^2)$ is still given by the naive $SU(6)$-symmetry expectation, $- 3 /
2$. However, it is also shown that the full Melosh rotations break $SU(6)$
symmetry, giving rise to $G_E^n(Q^2) \neq 0$ and $G_M^p(Q^2) / G_M^n(Q^2)
\neq - 3 / 2$ even when a $SU(6)$-symmetric canonical wave function is
assumed. It turns out that relativistic effects alone cannot explain
simultaneously the experimental data on $G_E^n(Q^2)$ and $G_M^p(Q^2) /
G_M^n(Q^2)$.   

\end{abstract}

\newpage

\pagestyle{plain}

\indent The elastic nucleon form factors contain important pieces of
information on the internal structure of the nucleon and therfore an
extensive program aimed at their experimental investigation is currently
undergoing and planned at several existing facilities around the world
\cite{Petratos}. In what follows we will focus on the so-called nucleon
Sachs form factors, which are defined as \cite{Sachs}
 \be
    G_E^N(Q^2) & = & F_1^N(Q^2) - {Q^2 \over 4M^2} F_2^N(Q^2) \nonumber \\
    G_M^N(Q^2) & = & F_1^N(Q^2) + F_2^N(Q^2)
    \label{eq:sachs}
 \ee
where $F_1^N(Q^2)$ [$F_2^N(Q^2)$] is the Dirac [Pauli] nucleon form factor,
appearing in the usual covariant decomposition of the nucleon electromagnetic
current matrix elements, viz.
 \be
    <N(p',s')| j_{em}^{\mu}(0) |N(p, s)> = \bar{u}(p', s') \left \{ 
    F_1^N(Q^2) \gamma^{\mu} + F_2^N(Q^2) {i \sigma^{\mu \nu} q_{\nu} \over 
    2M} \right \} u(p, s)
    \label{eq:F12}
 \ee     
with $Q^2 = - q \cdot q$ and $M$ being the squared four-momentum transfer and
the nucleon mass, respectively. As it is well known, the nucleon Sachs form
factors may be interpreted in the Breit frame as the Fourier transforms of
the nucleon charge and magnetisation density, respectively. In this 
respect, from Eq. (\ref{eq:sachs}) the squared nucleon charge radius, 
$r_N^2$, is therefore given by the sum of two terms, namely
 \be
    r_N^2 \equiv - 6 \left[{dG_E^N(Q^2) \over dQ^2} \right]_{Q^2 = 0} = 
    r_{1N}^2 + {3 k_N \over 2M^2}
    \label{eq:radius}
 \ee
where $r_{1N}^2 \equiv - 6 [dF_1^N(Q^2) / dQ^2]_{Q^2 = 0}$ and $k_N \equiv
F_2^N(0)$ is the nucleon anomalous magnetic moment. The second term in the 
r.h.s of Eq. (\ref{eq:radius}) is usually referred to as the Foldy
contribution and it is of relativistic origin.

\indent In case of the neutron the charge radius has been nicely determined
from electron-neutron elastic scattering experiments at very low energy, 
obtaining $r_n^2 = -0.113 \pm 0.005 ~ fm^2$ \cite{radius_n}. Since from Eq. 
(\ref{eq:radius}) the Foldy contribution turns out to be $\simeq -0.126 ~ 
fm^2$, the experimental value of the neutron charge radius appears to be 
almost totally explained by its Foldy term alone, i.e. by relativistic 
effects. This result, which implies a small value for $r_{1N}^2$ ($\simeq 
0.013 \pm 0.005 ~ fm^2$), has been viewed \cite{Weise} as an indication of 
the smallness of the {\em intrinsic} charge radius related to the neutron
rest-frame charge distribution. Nevertheless, very recently \cite{Isgur} the 
interpretation of the neutron charge radius as arising from its internal 
charge distribution has been asserted again. Indeed, it has been argued 
that, going beyond the non-relativistic limit when the Foldy term firstly 
appears, the Dirac neutron form factor $F_1^n(Q^2)$ receives a relativistic 
correction that cancels exactly against the Foldy term in $G_E^n(Q^2)$. Such
a statement has been inferred from the observation that the well-known 
phenomenon of {\em zitterbewegung}, which produces the Foldy term, cannot 
contribute to the charge radius of the neutron, because the latter has zero
total charge \cite{Isgur}.

\indent The aim of this letter is to address the issue of the relativistic
effects on $G_E^n(Q^2)$ adopting a constituent quark ($CQ$) model formulated
on the light-front.

\indent Let us briefly recall the basic notation and the relevant structure
of the nucleon wave function in the light-front formalism. Following, e.g.,
Refs. \cite{Coester,CAR95,CAR99} the light-front nucleon wave function is 
eigenstate of the non-interacting angular momentum operators $j^2$ and
$j_n$, where the vector $\hat{n} = (0, 0, 1)$ defines the spin quantization
axis. For a system of three quarks with equal masses the squared free-mass
operator is given by  $M_0^2 = \sum_{i = 1}^3 (k_{i \perp}^2 + m^2) /
\xi_i$, where $m$ is the $CQ$ mass, $\xi_i = p_i^+ / P^+$ and $\vec{k}_{i
\perp} = \vec{p}_{i \perp} - \xi_i \vec{P}_{\perp}$ are the intrinsic
light-front variables. The subscript $\perp$ indicates the projection
perpendicular to the spin quantization axis and the {\em plus} component of
a 4-vector $p \equiv (p^0, \vec{p})$ is given by $p^+ = p^0 + \hat{n} \cdot
\vec{p}$; finally $\tilde{P} \equiv (P^+, \vec{P}_{\perp}) = \tilde{p}_1 +
\tilde{p}_2 + \tilde{p}_3$ is the light-front nucleon momentum and
$\tilde{p}_i$ the quark one. In terms of the longitudinal momentum $k_{in}$,
related to the variable $\xi_i$ by $k_{in} = \left[ \xi_i M_0 - (k_{i
\perp}^2 + m^2) / \xi_i M_0 \right] / 2$,  the free mass operator acquires a
familiar form, viz. $M_0 = \sum_{i = 1}^3 \sqrt{m^2 + k_i^2} = \sum_{i =
1}^3 E_i$  with $\vec{k}_i \equiv ( \vec{k}_{i \perp}, k_{in})$.
Disregarding the colour degrees of freedom, the light-front nucleon wave
function can be written as
 \be
    \langle \{ \xi_i \vec{k}_{i \perp}; \nu'_i \tau_i \}|
    \Psi_{N}^{\nu_N}\rangle ~ = ~ \sqrt{{E_1 E_2 E_3 \over M_0 \xi_1 \xi_2 
    \xi_3}} ~ \sum_{ \{\nu_i \}} ~ \langle \{ \nu'_i \} |
    {\cal{R}}^{\dag}|\{\nu_i \} \rangle \langle \{\vec{k_i}; \nu_i \tau_i \} 
    | \chi_{N}^{\nu_N} \rangle 
    \label{eq:wfLF} 
 \ee 
where the curly braces $\{ ~~ \}$ mean a list of indices corresponding to $i
= 1, 2, 3$; $\nu_i$ ($\tau_i$) is the third component of the quark spin
(isospin); ${\cal{R}}^{\dag} = \prod_{j = 1}^3  R_j^{\dag}(\vec{k}_{j \perp},
\xi_j, m)$ is the product of individual  (generalised) Melosh  rotations,
viz.
 \be
    R_j(\vec{k}_{j \perp}, \xi_j, m) \equiv {m + \xi_j M_0 - i 
    \vec{\sigma}^{(j)} \cdot (\hat{n} \times \vec{k}_{j \perp}) \over 
    \sqrt{(m + \xi_j M_0)^2 + k_{j \perp}^2}}
    \label{eq:melosh}
 \ee
with $\vec{\sigma}$ being the ordinary Pauli spin matrices. In what follows
we will limit ourselves to the case of a pure $SU(6)$ symmetric canonical (or
equal-time) wave function, namely
 \be
    \langle \{ \vec{k}_i; \nu_i \tau_i\}| \chi_{N}^{\nu_N} \rangle = ~
    w_S(\vec{k}, \vec{p}) ~ {1 \over \sqrt{2}}~ \left[ \Phi^{00}_{\nu_N 
    \tau_N}(\{\nu_i \tau_i \}) + \Phi^{11}_{\nu_N \tau_N}(\{ \nu_i 
    \tau_i \}) \right]
    \label{eq:canonical}
 \ee
where $\vec{k} = (\vec{k}_1 - \vec{k}_2) / 2$ and $\vec{p} = \vec{k}_3$ are
the Jacobian internal co-ordinates for the three-quark system, and 
$w_S(\vec{k}, \vec{p})$ is a completely symmetric radial $S$-wave function.
Finally, the spin-isospin function $\Phi^{S_{12} T_{12}}_{\nu_N \tau_N}(\{
\nu_i\tau_i \})$ is defined as
 \be
    \Phi^{S_{12} T_{12}}_{\nu_N\tau_N}(\{ \nu_i\tau_i \}) & = & \sum_{M_S} 
    \langle {1 \over 2} \nu_1 {1 \over 2} \nu_2 | S_{12} M_S \rangle ~ 
    \langle S_{12} M_S {1 \over 2} \nu_3 | {1 \over 2} \nu_N \rangle \cdot
    \nonumber \\
    & & \sum_{M_T} \langle {1 \over 2} \tau_1 {1 \over 2} \tau_2 | T_{12} 
    M_T \rangle ~ \langle T_{12} M_T {1 \over 2} \tau_3 | {1 \over 2} 
    \tau_N \rangle 
    \label{eq:STwf}
 \ee
where $S_{12}$ ($T_{12}$) is the total spin (isospin) of the quark pair $(1,
2)$. The normalisation of the wave function (\ref{eq:canonical}) is: 
$\sum_{\{\nu_i \tau_i \}} \int d\vec{k}_1 ~ d\vec{k}_2 ~ d\vec{k}_3 ~
\delta(\vec{k}_1 + \vec{k}_2 + \vec{k}_3) \left| \langle \{ \vec{k}_i, \nu_i
\tau_i \} | \chi_{N}^{\nu_N} \rangle \right|^2 = \int d\vec{k} ~ d\vec{p}
\left| w_S(\vec{k}, \vec{p}) \right|^2$ $= 1$.

\indent As in Refs. \cite{CAR95,CAR99} we consider the one-body component of
the electromagnetic (e.m.) current operator including $CQ$ form factors,
namely 
 \be 
    {\cal{I}}^{\nu} = \sum_{j = 1}^3 I_j^{\nu} = \sum_{j=1}^3 \left[ e_j
    \gamma^{\nu} f_1^j(Q^2) ~ + ~ i \kappa_j {\sigma^{\nu \mu} q_{\mu}
    \over 2m} f_2^j(Q^2) \right]
    \label{eq:current}
 \ee
where $\sigma^{\nu \mu} = {i \over 2}[\gamma^{\nu},\gamma^{\mu}]$, $e_j$ is
the charge of the j-th quark, $\kappa_j$ the corresponding anomalous magnetic
moment and $f_{1(2)}^j(Q^2)$ its Dirac (Pauli) form factor (with
$f_{1(2)}^j(0) = 1$). It is  well known that in the light-front formalism
the form factors for a  conserved current can be obtained using only the
matrix elements of the  {\em plus} component of the current operator (see,
e.g., Ref. \cite{Coester})  and, moreover, for $Q^2 \geq 0$ the choice of a
frame where $q^+ = 0$  allows to suppress the contribution of the $Z$-graph
(i.e., pair creation  from the vacuum) \cite{Zgraph}. More precisely, in
what follows the  four-momentum transfer $q$ is taken to be purely
transverse, i.e. $q =  (0,\vec{q}_{\perp},0)$ with $Q^2 = -q^2 =
|\vec{q}_{\perp}|^2$. In case of the nucleon one has
 \be 
    \langle \Psi_{N}^{\nu'_N}| ~ {\cal{I}}^+ ~ |\Psi_N^{\nu_N} \rangle =
    F_1^N(Q^2) ~ \delta_{{\nu'}_N \nu_{N}} - i {Q \over 2 M} 
    F_2^N(Q^2) ~ \langle \nu'_{N} | \sigma_y | \nu_{N} \rangle
    \label{eq:Iplus}
 \ee
and therefore $F_1^N(Q^2)$ is related to the non spin-flip matrix elements,
while $F_2^N(Q^2)$ can be extracted from the spin-flip ones. Using Eqs. 
(\ref{eq:wfLF}-\ref{eq:Iplus}) the nucleon Dirac and Pauli form factors  are
explicitly given by \cite{fabio}
 \be
    F_1^{p(n)}(Q^2) & = & {3 \over 2} \int [d\xi] ~ [d\vec{k}_{\perp}] ~ 
    [d\vec{k'}_{\perp}] ~ \sqrt{ {E_1 E_2 E_3 {M'}_0 \over {E'}_1 {E'}_2 
    {E'}_3 M_0}} ~ w_S^*(\vec{k'}, \vec{p'}) ~ w_S(\vec{k}, \vec{p}) ~ 
    \left( \prod_{j = 1}^3 {\cal{N'}}_j {\cal{N}}_j \right) \cdot
    \nonumber \\
    & & \left \{ \left[ A_1 A_2 + \vec{B}_1 \cdot \vec{B}_2 \right] ~ 
    \left[ A_3 e_{u(d)} f_1^{u(d)}(Q^2) + \tilde{A}_3 {Q \over 2m} 
    \kappa_{u(d)} f_2 ^{u(d)}(Q^2) \right] ~ + \right.
    \nonumber \\
    & & \left. 
    {1 \over 9} \left[ \left( 3A_1 A_2 - \vec{B}_1 \cdot \vec{B}_2 \right) 
    ~ A_3  + 2 \left( A_1 \vec{B}_2 + A_2 \vec{B}_1 \right) \cdot \vec{B}_3 
    \right] \cdot \right. 
    \nonumber \\
    & & \left. \left[ e_{u(d)} f_1^{u(d)}(Q^2) + 2 e_{d(u)} f_1^{d(u)}(Q^2) 
    \right] + {1 \over 9} \left[ \left( 3A_1 A_2 - \vec{B}_1 \cdot \vec{B}_2 
    \right) ~ \tilde{A}_3 + \right. \right.
    \nonumber \\
    & & \left. \left. 2 \left( A_1 \vec{B}_2 + A_2 \vec{B}_1 \right) \cdot 
    \vec{\tilde{B}}_3 \right] {Q \over 2m} \left[ \kappa_{u(d)} 
    f_2^{u(d)}(Q^2) + 2 \kappa_{d(u)} f_2^{d(u)}(Q^2) \right] \right \}
    \label{eq:F1}
 \ee
 \be
    F_2^{p(n)}(Q^2) & = & - {3M \over Q} \int [d\xi] ~ [d\vec{k}_{\perp}] ~ 
    [d\vec{k'}_{\perp}] ~ \sqrt{ {E_1 E_2 E_3 {M'}_0 \over {E'}_1 {E'}_2 
    {E'}_3 M_0}} ~ w_S^*(\vec{k'}, \vec{p'}) ~ w_S(\vec{k}, \vec{p}) ~ 
    \left( \prod_{j = 1}^3 {\cal{N'}}_j {\cal{N}}_j \right) \cdot
    \nonumber \\
    & & \left \{ \left[ A_1 A_2 + \vec{B}_1 \cdot \vec{B}_2 \right] ~ 
    \left[ B_{3y} e_{u(d)} f_1^{u(d)}(Q^2) + \tilde{B}_{3y} {Q \over 2m} 
    \kappa_{u(d)} f_2 ^{u(d)}(Q^2) \right] ~ + \right.
    \nonumber \\
    & & \left. {1 \over 9} \left[ 2 \left( B_{1y} \vec{B}_2 + B_{2y} 
    \vec{B}_1 \right) \cdot \vec{B}_3 - B_{3y} \left( A_1 A_2 + \vec{B}_1 
    \cdot \vec{B}_2 \right) ~ + \right. \right.
    \nonumber \\
    & & \left. \left. 2 A_3 \left( A_1 B_{2y} + A_2 B_{1y} \right) \right] 
    \left[ e_{u(d)} f_1^{u(d)}(Q^2) + 2 e_{d(u)} f_1^{d(u)}(Q^2) \right] ~ 
    + \right.
    \nonumber \\
    & & \left. {1 \over 9} \left[ 2 \left( B_{1y} \vec{B}_2 + B_{2y} 
    \vec{B}_1 \right) \cdot \vec{\tilde{B}}_3 - \tilde{B}_{3y} \left( 
    A_1 A_2 + \vec{B}_1 \cdot \vec{B}_2 \right) ~ + \right. \right. 
    \nonumber \\
    & & \left. \left. 2 \tilde{A}_3 \left( A_1 B_{2y} + A_2 B_{1y} \right) 
    \right] {Q \over 2m} \left[ \kappa_{u(d)} f_2^{u(d)}(Q^2) + 2 
    \kappa_{d(u)} f_2^{d(u)}(Q^2) \right] \right \}
    \label{eq:F2}
 \ee
where
 \be
    \left[ d\xi \right] & = & d\xi_1 ~ d\xi_2 ~ d\xi_3 ~ \delta[\xi_1 + 
    \xi_2 + \xi_3 - 1]
    \nonumber \\
    \left[ d\vec{k}_{\perp} \right] & = & d\vec{k}_{1 \perp} ~ 
    d\vec{k}_{2 \perp} ~ d\vec{k}_{3 \perp} ~ \delta[\vec{k}_{1 \perp} + 
    \vec{k}_{2 \perp} + \vec{k}_{3 \perp}]
    \nonumber \\
    \left[ d\vec{k'}_{\perp} \right] & = & d\vec{k'}_{1 \perp} ~ d\vec{k'}_{2 
    \perp} ~ d\vec{k'}_{3 \perp} ~ \delta[\vec{k'}_{1 \perp} - \vec{k}_{1 
    \perp} + \xi_1 \vec{q}_{\perp}] ~ \delta[\vec{k'}_{2 \perp} - \vec{k}_{2 
    \perp} + \xi_2 \vec{q}_{\perp}] 
    \nonumber \\
    & & \delta[\vec{k'}_{3 \perp} - \vec{k}_{3 \perp} + (\xi_3 - 1 ) ~ 
    \vec{q}_{\perp}]
    \label{eq:int}
 \ee
and
 \be
    {\cal{N}}_j & = & 1 / \sqrt{(m + \xi_j M_0)^2 + k_{j \perp}^2}
    \nonumber \\
    {\cal{N'}}_j & = & 1 / \sqrt{(m + \xi_j {M'}_0)^2 + {k'}_{j \perp}^2}
    \nonumber \\
    A_j & = & (m + \xi_j {M'}_0) ~ (m + \xi_j M_0) + \vec{k'}_{j \perp} 
    \cdot \vec{k}_{j \perp}
    \nonumber \\
    \tilde{A}_3 & = & (m + \xi_3 {M'}_0) k_{3x} - (m + \xi_3 M_0) {k'}_{3x}
    \nonumber \\ 
    B_{jx} & = & (m + \xi_j M_0) {k'}_{jy} - (m + \xi_j {M'}_0) k_{jy}
    \nonumber \\
    B_{jy} & = & (m + \xi_j {M'}_0) k_{jx} - (m + \xi_j M_0) {k'}_{jx}
    \nonumber \\
    B_{jz} & = & {k'}_{jx} k_{jy} - {k'}_{jy} k_{jx}
    \nonumber \\
    \tilde{B}_{3x} & = & {k'}_{3x} k_{3y} + {k'}_{3y} k_{3x}
    \nonumber \\
    \tilde{B}_{3y} & = & - (m + \xi_3 {M'}_0) (m + \xi_3 M_0) - {k'}_{3x} 
    k_{3x} + {k'}_{3y} k_{3y}
    \nonumber \\
    \tilde{B}_{3z} & = & (m + \xi_3 {M'}_0) k_{3y} + (m + \xi_3 M_0) 
    {k'}_{3y}
    \label{eq:ABN}
 \ee

\indent In the non-relativistic limit, assuming both point-like constituents
and the SU(6)-symmetric wave function (\ref{eq:canonical}), the Fourier
transform of the non-relativistic charge density, $\sum_{j = 1}^3 e_j ~ 
\delta(\vec{r} - \vec{r_j})$, yields $G_E^N(Q^2) \to \sum_{j = 1}^3 e_j 
F_c(Q^2)$, i.e. $G_E^p(Q^2) \to F_c(Q^2)$ and $G_E^n(Q^2) \to 0$, with 
 \be
    F_c(Q^2 = |\vec{q}|^2) \equiv \int d\vec{k} ~ d\vec{p} ~ w_S^*(\vec{k}, 
    \vec{p} + {2 \over 3} \vec{q}) ~ w_S(\vec{k}, \vec{p})
    \label{eq:Fc}
 \ee
In a similar way, in case of the non-relativistic magnetisation density the
nucleon magnetic form factors are given by $G_M^p(Q^2) \to 3 F_c(Q^2)$ and
$G_M^n(Q^2) \to -2 F_c(Q^2)$. Note that the $SU(6)$ symmetry predicts both
$G_E^n(Q^2) = 0$ and $G_M^p(Q^2) / G_M^n(Q^2) = - 3 / 2$.

\indent To reach the non-relativistic limit from Eqs.
(\ref{eq:F1}-\ref{eq:F2}) is not a trivial task; moreover, as argued in Ref.
\cite{Isgur}, a subtle cancellation of relativistic corrections is expected
to hold when the Foldy term firstly appears. We now want to find out the
assumptions that allow to obtain the specific cancellation claimed in Ref.
\cite{Isgur} and, then, we want to carry out the appropriate
non-relativistic reduction of Eqs. (\ref{eq:F1}-\ref{eq:F2}). The basic
assumption made in Ref. \cite{Isgur} is to neglect the transverse motion of
the constituents in order to avoid spin-flip effects from the Wigner
rotations of the constituent spins. Thus, our first  assumption is to put
$\vec{k}_{j \perp} = 0$ in the Melosh rotations (\ref{eq:melosh}), yielding
for $j = 1, 2, 3$
 \be
    R_j(\vec{k}_{j \perp}, \xi_j, m) \to 1
    \label{eq:hyp1}
 \ee
As for the Melosh rotations in the final state (i.e., after virtual photon
absorption), we cannot make the same assumption (\ref{eq:hyp1}), otherwise
in case of point-like constituents only a vanishing $F_2^N(Q^2)$ could be 
obtained\footnote{A vanishing $F_2^N(Q^2)$ is appropriate in the case of the
heavy-quark limit $m \to \infty$ (cf. Ref. \cite{CAR98}).}. Thus, inspired
by Ref. \cite{Coester}, we consider the following approximation for $j = 1,
2, 3$
 \be
    R_j(\vec{k'}_{j \perp}, \xi_j, m) \to {2m - i \vec{\sigma}^{(j)} 
    \cdot (\hat{n} \times \vec{k'}_{j \perp}) \over \sqrt{4m^2 + 
    {k'}_{j \perp}^2}}
    \label{eq:hyp2}
 \ee
where the denominator is included to maintain the correct normalisation of
the final light-front wave function. Note that the term proportional to
$\vec{\sigma}^{(j)}$ in the r.h.s. of Eq. (\ref{eq:hyp2}) is clearly
reminiscent of the non-relativistic magnetisation current. Finally, we have
to specify the final transverse momenta $\vec{k'}_{j \perp}$ appearing in
Eq. (\ref{eq:hyp2}). To this end we neglect the initial longitudinal motion,
i.e. we consider  $k_{jn} = 0$, which together with the assumption
$\vec{k}_{j \perp} = 0$  yields $\xi_j = 1/3$. Thus, from the delta
functions in Eq. (\ref{eq:int}) the final transverse momenta turn out to be:
$\vec{k'}_{1 \perp} =  \vec{k'}_{2 \perp} = - \vec{q}_{\perp} / 3$ and
$\vec{k'}_{3 \perp} = 2  \vec{q}_{\perp} / 3$. Our assumptions
(\ref{eq:hyp1}-\ref{eq:hyp2})  correspond to consider $\xi_j M_0 = \xi_j
{M'}_0 = m$ in Eq. (\ref{eq:ABN}), leading to: ${\cal{N}}_j = 1 / 2m$,
${\cal{N'}}_1 = {\cal{N'}}_2 = 1 / \sqrt{4m^2 + Q^2 / 9}$, ${\cal{N'}}_3 = 1
/ \sqrt{4m^2 + 4Q^2 / 9}$, $A_j = 4m^2$, $\tilde{A}_3 = 2mQ / 3$, $B_{1y} =
B_{2y} = 2mQ /3$, $B_{3y} = - 4mQ / 3$, $\tilde{B}_{3y} = - 4m^2$, while all
the other  $B$'s and $\tilde{B}$'s are identically vanishing. Finally, for
compatibility with the non-relativistic reductions $G_M^p(0) \to 3$ and 
$G_M^n(0) \to - 2$ we neglect any binding effect in the nucleon mass, i.e.
we consider $M = 3m$ (cf. also Refs. \cite{Isgur,Coester}).   

\indent Before carrying out the effects of the assumptions
(\ref{eq:hyp1}-\ref{eq:hyp2}) on Eqs. (\ref{eq:F1}-\ref{eq:F2}), we have  to
restrict ourselves to the case of $SU(2)$-symmetric $CQ$ form factors, 
otherwise any deviation of $G_E^n(Q^2)$ from zero could be attributed to  a
possible flavour-dependence of the $CQ$ internal composite structure. Thus,
in what follows we will consider
 \be
    f_1^q(Q^2) & = & f(Q^2) \nonumber \\
    k_q f_2^q(Q^2) & = & e_q \kappa \tilde{f}(Q^2)
    \label{eq:CQff}
 \ee
where $f(Q^2)$, $\tilde{f}(Q^2)$ and $\kappa$ do not depend on the flavour of
the $CQ$ (and $f(0) = \tilde{f}(0) = 1$). Using all the above-discussed 
assumptions it is straightforward to obtain from Eqs. 
(\ref{eq:F1}-\ref{eq:F2}) the following explicit expressions
 \be
    F_1^p(Q^2) & = & F_0(Q^2) + 3\kappa {Q^2 \over 4M^2} \tilde{F}_0(Q^2)
    \nonumber \\
    F_1^n(Q^2) & = & -2F_0(Q^2) {Q^2 \over 4M^2 + Q^2} - 2\kappa 
    \tilde{F}_0(Q^2) {Q^2 \over 4M^2}
    \nonumber \\
    F_2^p(Q^2) & = & 2F_0(Q^2) + 3\kappa \tilde{F}_0(Q^2)
    \nonumber \\
    F_2^n(Q^2) & = & -2F_0(Q^2) {4M^2 \over 4M^2 + Q^2} - 2\kappa 
    \tilde{F}_0(Q^2)
    \label{eq:F1F2}
 \ee
where
 \be
    F_0(Q^2) \equiv {f(Q^2) \over \sqrt{1 + Q^2 / M^2}} \int [d\xi] ~ 
    [d\vec{k}_{\perp}] ~ [d\vec{k'}_{\perp}] ~ \sqrt{ {E_1 E_2 E_3 {M'}_0
    \over {E'}_1 {E'}_2 {E'}_3 M_0}} ~ w_S^*(\vec{k'}, \vec{p'}) ~ 
    w_S(\vec{k}, \vec{p})
    \label{eq:F0}
 \ee
while $\tilde{F}_0(Q^2)$ is given by Eq. (\ref{eq:F0}) but with $f(Q^2)$
replaced by $\tilde{f}(Q^2)$. In terms of the Sachs form factors
(\ref{eq:sachs}) one gets
 \be
    G_E^p(Q^2) & = & (1 - {Q^2 \over 2M^2}) F_0(Q^2)
    \nonumber \\
    G_E^n(Q^2) & = & 0
    \nonumber \\
    G_M^p(Q^2) & = & 3 \left[ F_0(Q^2) + \kappa (1 + {Q^2 \over 4M^2}) 
    \tilde{F}_0(Q^2)\right]
    \nonumber \\
    G_M^n(Q^2) & = & - 2 \left[ F_0(Q^2) + \kappa (1 + {Q^2 \over 4M^2}) 
    \tilde{F}_0(Q^2)\right]
    \label{eq:GEGM}
 \ee

\indent The relativistic corrections to $F_{1,2}^N(Q^2)$, contained in Eq.
(\ref{eq:F1F2}), largely differ for proton and neutron, but: i) $F_1^n(Q^2)$
receives a relativistic correction that cancels exactly against the Foldy
term, $- F_2^n(Q^2) \cdot Q^2 / 4M^2$, so that the $SU(6)$-symmetry
prediction $G_E^n(Q^2) = 0$ still holds, as argued in Ref. \cite{Isgur}; the
same type of cancellation does not occur in case of the proton charge form
factor $G_E^p(Q^2)$; ii) the ratio $G_M^p(Q^2) / G_M^n(Q^2)$ is still given
by the simple $SU(6)$-symmetry expectation, $- 3 / 2$. Both results are
independent of any particular choice of the $CQ$ form factors (both Dirac
and Pauli ones), provided the latter are taken to be $SU(2)$ symmetric (see
Eq. (\ref{eq:CQff})). Note that Eq. (\ref{eq:GEGM}) predicts a vanishing
value for $G_E^p(Q^2)$ at $Q^2 =  2M^2$; this is completely at variance with
experimental data and signals that the applicability of Eqs. (\ref{eq:F1F2})
and (\ref{eq:GEGM}) is limited only to low values of $Q^2$ (i.e., $Q^2 <<
M^2$). Finally, note also that in Ref. \cite{Coester} it is claimed that at
first order in $Q^2 / M^2$ the leading term in $G_E^n(Q^2)$ is given by its
Foldy term. Such a statement is incorrect, because it is obtained using
directly in $G_E^n(Q^2)$ the non-relativistic limits of $F_1^n$ and
$F_2^n(Q^2)$, ignoring in this way the Isgur's cancellation mechanism of the
Foldy term.

\indent We can now carry out the non-relativistic limit of the nucleon form
factors by considering the formal limit $M \to \infty$ in Eqs.
(\ref{eq:F1F2}) and (\ref{eq:GEGM}). Taking also into account that the
non-relativistic reduction of the integral in Eq. (\ref{eq:F0}) leads to Eq.
(\ref{eq:Fc}) (see Ref. \cite{Coester}), one has
 \be
    F_1^p(Q^2) & \to & F_c(Q^2) ~ f(Q^2)
    \nonumber \\
    F_1^n(Q^2) & \to & 0
    \nonumber \\
    F_2^p(Q^2) & \to & F_c(Q^2) ~ [2f(Q^2) + 3\kappa \tilde{f}(Q^2)]
    \nonumber \\
    F_2^n(Q^2) & \to & - 2F_c(Q^2) ~ [f(Q^2) + \kappa \tilde{f}(Q^2)]
    \label{eq:F1F2_NR}
 \ee
and
 \be
    G_E^p(Q^2) & \to & F_c(Q^2) ~ f(Q^2)
    \nonumber \\
    G_E^n(Q^2) & \to & 0
    \nonumber \\
    G_M^p(Q^2) & \to & 3F_c(Q^2) ~ [f(Q^2) + \kappa \tilde{f}(Q^2)]
    \nonumber \\
    G_M^n(Q^2) & \to & - 2F_c(Q^2) ~ [f(Q^2) + \kappa \tilde{f}(Q^2)]
    \label{eq:GEGM_NR}
 \ee

\indent It is now worthwhile to make the following two observations:

\begin{itemize}

\item{the approximation leading to Eqs. (\ref{eq:F1F2}) and (\ref{eq:GEGM})
is mainly based on neglecting the constituent initial transverse motion in
the Melosh rotations. Thus, Eq. (\ref{eq:F1F2}) can represent a good
approximation of the full calculations (\ref{eq:F1}-\ref{eq:F2}) only when
the average value of the transverse momenta, $\langle p_{\perp} \rangle$, is
much smaller than the constituent mass $m$. However, in $QCD$ both $m$ and
$\langle p_{\perp} \rangle$ are expected to be of the order of the $QCD$
scale, $\Lambda_{QCD} \sim 300 ~ MeV$. Moreover, in quark potential models
$\langle p_{\perp} \rangle$ turns out to be significantly larger than $m$,
because of the high momentum components generated in the light-baryon wave
functions by the short-range part of the effective quark-quark interaction
\cite{CAR99};}

\item{the Melosh rotations break in general the $SU(6)$ symmetry. Indeed,
these rotations, being momentum and spin dependent, produce a re-coupling of
the constituent orbital angular momentum and spin; in other words, even if
the canonical wave function (\ref{eq:canonical}) is assumed to be factorized
into a spatial part times a spin-isospin wave function (which can be
classified according to the $SU(6)$ multiplets), after the application of
the Melosh rotations (\ref{eq:melosh}) the light-front wave function
(\ref{eq:wfLF}) cannot be any more expressed as a product of a spatial part
times a spin-isospin function. As a result, the light-front wave function
(\ref{eq:wfLF}) is not $SU(6)$ symmetric and therefore we do not expect in
general to have $G_E^n(Q^2) = 0$ and $G_M^p(Q^2) / G_M^n(Q^2) = - 3 / 2$.}

\end{itemize}

\noindent Consequently, the inclusion of the effects of the quark initial
transverse motion in Eqs. (\ref{eq:F1}-\ref{eq:F2}) could lead to
$G_E^n(Q^2) \neq 0$ and $G_M^p(Q^2) / G_M^n(Q^2) \neq - 3 / 2$. The former
point is  clearly illustrated in Fig. 1, where the results of the
calculations of Eqs. (\ref{eq:F1}-\ref{eq:F2}), performed for point-like
$CQ$'s adopting a simple gaussian-like ans\"atz for the radial function
$w_S(\vec{k}, \vec{p}) \propto exp[-(k^2 + 3p^2 / 4) / 2 a_{HO}^2]$, are
reported for various values of the quantity $\langle p_{\perp} \rangle =
\sqrt{4/3} ~ a_{HO}$ at $m = 220 ~ MeV$ (chosen as in Ref. \cite{CI86}). It
can clearly be seen that relativistic effects may contribute significantly
to $G_E^n(Q^2)$ when $\langle p_{\perp} \rangle > m$\footnote{This result
does not depend on the particular choice of a gaussian-like ans\"atz for the
radial function $w_S(\vec{k}, \vec{p})$, provided the average value $\langle
p_{\perp} \rangle$ is kept the same. In particular, the nucleon
eigenfunction corresponding to the quark potential model of Ref. 
\cite{CI86}, yields $\langle p_{\perp} \rangle \simeq 0.58 ~ GeV$;
we have checked that the use of the $SU(6)$-symmetric part of this wave
function leads to results almost coinciding with the solid curve of Fig. 1,
which corresponds indeed to  $\langle p_{\perp} \rangle \simeq 0.58 ~
GeV$.}. Note that, as $\langle p_{\perp} \rangle$ increases, the neutron
charge radius appears not to exceed $\sim 40 \%$ of its experimental value.

\indent The $SU(6)$ breaking associated to the Melosh rotations heavily
affects also the ratio $G_M^p(Q^2) / G_M^n(Q^2)$, as it is illustrated in
Fig. 2. It should be reminded that the experimental data on $G_M^p(Q^2)$ and
$G_M^n(Q^2)$ exhibit the well-known dipole behaviour, leading to $G_M^p(Q^2)
/ G_M^n(Q^2) \simeq \mu_p / \mu_n \simeq -1.46$ with only a $10 \div 15 \%$
uncertainty up to $Q^2 \sim 1 ~ (GeV/c)^2$ (cf., e.g., Ref.
\cite{Petratos}). Thus, Figs. 1 and 2 clearly indicate that relativistic
effects alone cannot explain simultaneously the experimental data on
$G_E^n(Q^2)$ and $G_M^p(Q^2) / G_M^n(Q^2)$; therefore, other mechanisms,
like, e.g., the presence of the mixed-symmetry S-wave generated by the
spin-spin forces among $CQ$'s and/or the effects of flavour-dependent $CQ$
form factors and/or the contribution of many-body e.m. currents, have to be
invoked.

\indent In conclusion, the neutron charge form factor $G_E^n(Q^2)$ has been
investigated within a constituent quark model formulated on the
light-front.  It has been shown that, if the quark initial motion is
neglected in the  Melosh rotations, the Dirac neutron form factor
$F_1^n(Q^2)$ receives a  relativistic correction which cancels exactly
against the Foldy term in $G_E^n(Q^2)$, as it has been recently argued in
Ref. \cite{Isgur}. The same type of cancellation does not occur in case of
the proton charge form factor $G_E^p(Q^2)$. Moreover, at the same level of
approximation the ratio  of the proton to neutron magnetic form factors
$G_M^p(Q^2) / G_M^n(Q^2)$ is still given by the naive $SU(6)$-symmetry
expectation, $- 3 / 2$. These results are independent of the electromagnetic
structure of the constituents, provided their form factors are $SU(2)$
symmetric. However, since the full Melosh rotations break $SU(6)$ symmetry,
both $G_E^n(Q^2) \neq 0$ and $G_M^p(Q^2) / G_M^n(Q^2) \neq - 3 / 2$ can be
obtained even when a $SU(6)$-symmetric canonical wave function is assumed.
It has been shown that relativistic effects alone cannot explain
simultaneously the experimental data on $G_E^n(Q^2)$ and $G_M^p(Q^2) /
G_M^n(Q^2)$.

\vspace{2cm}

\noindent{\bf Acknowledgements.} One of the authors, S.S., is deeply
indebted with Nathan Isgur for many fruitful discussions while at the
Institute for Nuclear Theory in Seattle, where part of the present work has
been carried out. The warm hospitality and the high-level scientific
environment found at the INT during the Int'l Workshop on {\em Algebraic
Methods in Many Body Physics}, organised by Franco Iachello and Joseph
Ginocchio, are also gratefully acknowledged.

\newpage

\begin{figure}[htb]

\centerline{\epsfxsize=16cm \epsfig{file=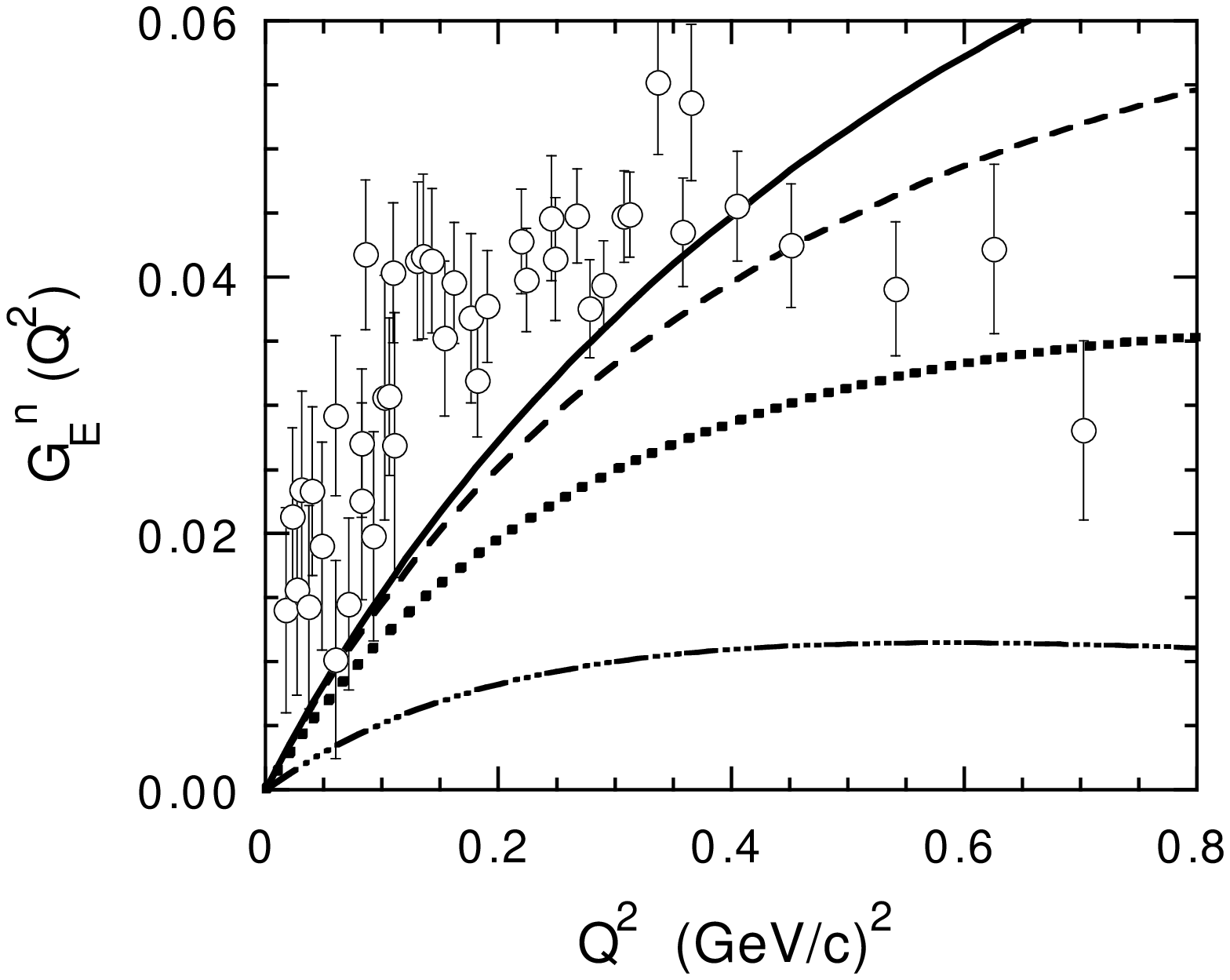}}

\rightline{} \vspace{1cm}

\small{{\bf Figure 1.} The neutron charge form factor $G_E^n(Q^2)$ versus
$Q^2$. The experimental data correspond to the results of the analysis of
Ref. \cite{Platchkov} performed in terms of the Reid-Soft-Core
nucleon-nucleon interaction. The various curves are the results of the
calculations of Eqs. (\ref{eq:F1}-\ref{eq:F2}), obtained assuming point-like
$CQ$'s and adopting a simple gaussian-like ans\"atz for the radial function
$w_S(\vec{k}, \vec{p})$ (see text). The value of the constituent quark mass
has been chosen to be $m = 220 ~ MeV$ from Ref. \cite{CI86}. The dot-dashed,
dotted, dashed and solid lines correspond to $\langle p_{\perp} \rangle =
\sqrt{4/3} ~ a_{HO} = 0.23, 0.35, 0.46$ and $0.58 ~ GeV$, respectively.}

\end{figure}

\newpage

\begin{figure}[htb]

\centerline{\epsfxsize=16cm \epsfig{file=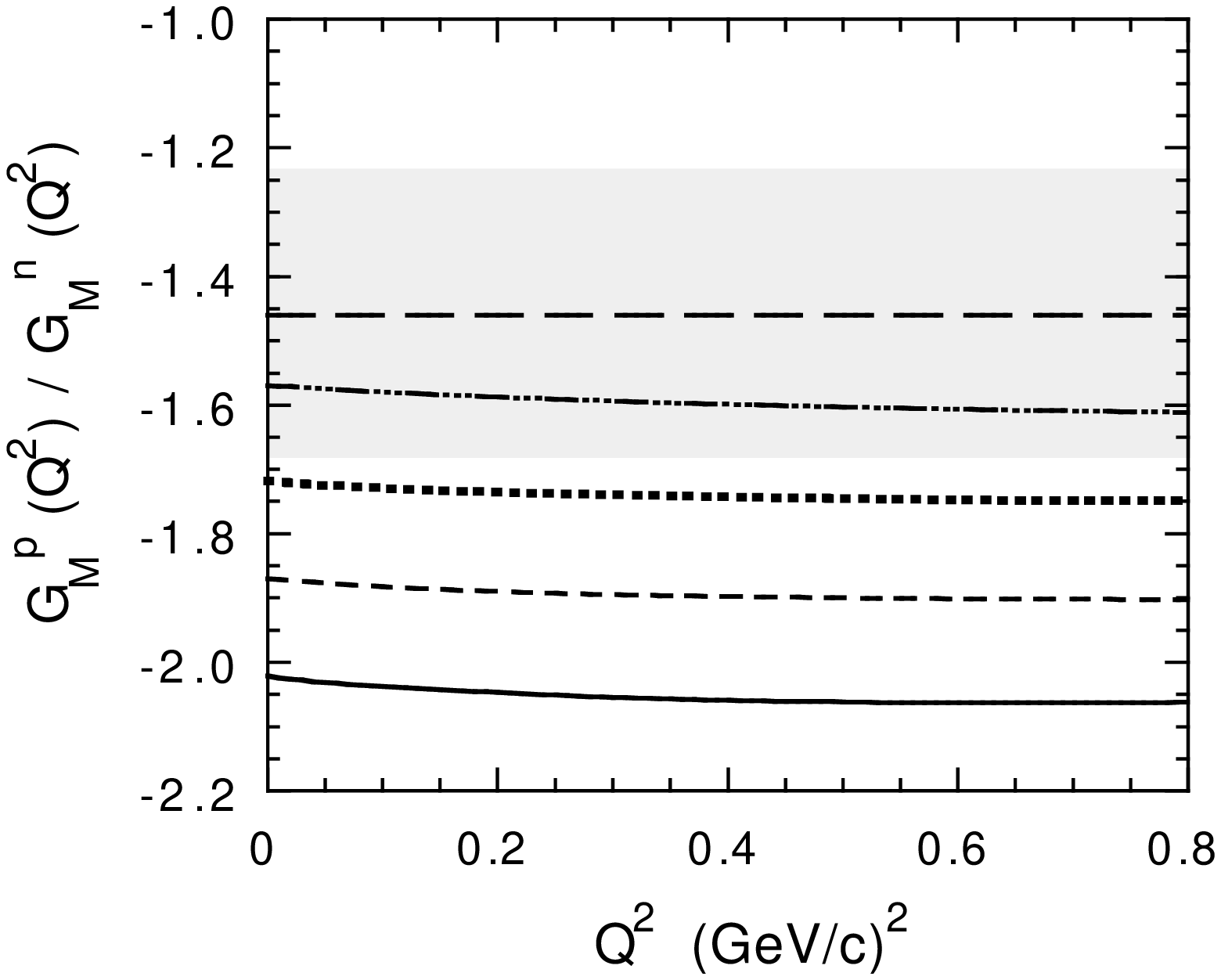}}

\rightline{} \vspace{1cm}

\small{{\bf Figure 2.} The ratio of the proton to neutron magnetic form
factors $G_M^p(Q^2) / G_M^n(Q^2)$ versus $Q^2$. The shaded area corresponds
to a $\pm 15 \%$ deviation from the dipole-fit expectation $G_M^p(Q^2) /
G_M^n(Q^2) \simeq \mu_p / \mu_n \simeq - 1.46$ (long-dashed line). The
dot-dashed, dotted, dashed and solid  lines are as in Fig. 1.}

\end{figure}


\newpage

\begin{thebibliography}{99}

\bibitem{Petratos} See, for a recent review, G. Petratos: in Proc. of the 
 Int'l Workshop on {\em The Structure of the Nucleon}, INFN National 
 Laboratory of Frascati (Italy), June 7-9, 1999, to appear in Nucl. Phys.
 A.

\bibitem{Sachs} R.G. Sachs: Phys. Rev. {\bf 126} (1962) 2256.

\bibitem{radius_n} S. Kopecki et al.: Phys. Rev. Lett. {\bf 74} (1995) 
 2427.

\bibitem{Weise} See, e.g., T. Ericson and W. Weise: {\em Pions and 
 Nuclei}, Claredon Press (Oxford, 1988).

\bibitem{Isgur} N. Isgur: Phys. Rev. Lett. {\bf 83} (1999) 272.

\bibitem{Coester} P.L. Chung and F. Coester: Phys. Rev. {\bf D44} (1991)
 229.

\bibitem{CAR95} F. Cardarelli et al.: Phys. Lett. {\bf B357} (1995) 267; 
 Few Body Syst. Suppl. {\bf 8} (1995) 345.

\bibitem{CAR99} F. Cardarelli et al.: Few Body Syst. Suppl. {\bf 11} 
 (1999) 66.

\bibitem{Zgraph} L.L. Frankfurt and M.I. Strikman: Nucl. Phys. {\bf B 148}
 (1979) 107. G.P. Lepage and S.J. Brodsky: Phys. Rev. {\bf D22} (1980)
 2157. S.J. Brodsky and G.P. Lepage: in {\em Perturbative Quantum
 Chromodynamics}, edited by A.H. Mueller, World Scientific (Singapore,
 1989), p. 93-240. T. Frederico and G.A. Miller: Phys. Rev. {\bf D45} 
 (1992) 4207. M. Sawicki: Phys. Rev. {\bf D46} (1992) 474.

\bibitem{fabio} F. Cardarelli: PhD thesis, University of Rome "La Sapienza",
 1995, unpublished.

\bibitem{CAR98} F. Cardarelli and S. Simula: Phys. Lett. {\bf B421} (1998) 
 295 and e-print archive hep-ph/9810414, to appear in Phys. Rev. D.

\bibitem{CI86} S. Capstick and N. Isgur: Phys. Rev. {\bf D34} (1986) 2809.

\bibitem{Platchkov} S. Platchkov et al.: Nucl. Phys. {\bf A510} (1990) 740.


\end{thebibliography}
\end{document}